\documentclass[a4paper,10pt]{article}
\usepackage{geometry,color,graphicx}
\usepackage{amssymb,amsmath}
\usepackage{graphicx}
\numberwithin{equation}{section}

\begin{document}
\title{\bf Thermodynamics and Thermodynamic geometry of Park black hole}
\author{{Jishnu Suresh \thanks{Email: jishnusuresh@cusat.ac.in} ,
\hspace{1mm} Tharanath R, \hspace{1mm} 
Nijo Varghese, \hspace{1mm}
and  V C Kuriakose} \\\\
\small{\em{Department of Physics, Cochin University of Science and Technology, Cochin 682022, Kerala, India}} } 

\date{}
\maketitle
\begin{abstract}
We study the thermodynamics and thermodynamic geometry of Park black hole in  Ho\v{r}ava gravity. By incorporating the ideas of differential
geometry, we have investigated the thermodynamics using Weinhold geometry and Ruppeiner geometry. We have also analyzed it in the context of
newly developed geometrothermodynamics(GTD). Divergence of specific heat is associated with the second order phase transition of black hole. Here in
the context of Park black hole, both Weinhold's metric and Ruppeiner's metric well explain this phase transition. But these explanations depend on 
the choice of potential. Hence the Legendre invariant GTD is used, and with the true singularities in the curvature scalar, GTD well explain the second
order phase transition. All these methods together give an exact idea of all the behaviors of the Park black hole thermodynamics.
\end{abstract}

\vspace{2cm}
\section{Introduction}

Over the past decade, a lot of interest has been given to various black holes in anti-de Sitter(AdS) space as well as in de-Sitter (dS) space, due to the 
success of AdS/CFT correspondence \cite{Maldacena} and hence the proposal of dS/CFT correspondence \cite{Witten,Strominger1,Strominger2}. So by 
studying the thermodynamics of these black holes, one may obtain a real connection between gravity and quantum mechanics. 
Ho\v{r}ava proposed \cite{horava1,horava2,horava3} a field theoretical model in 2009, which can be considered as a UV complete theory of gravity without
full diffeomorphism invariance. It can be reduced to Einstein's theory in the IR regime and is non-relativistic in the UV regime. Recently its 
black hole solutions and thermodynamics have been intensively investigated
\cite{LMP,Caicaoohta,KS,Nastase,Kofinas,Calcagni,Park1,Wei,Myung,Myung1,Kim,nv1,js1} .
Later, Park has obtained a $\lambda=1$ black solution, known
as Park black hole \cite{Park}. By introducing two parameters $\omega$ and $\Lambda_W$, Park found that both dS and AdS solutions exist.

In almost all macroscopic systems, usual thermodynamics entirely depends on the empirical
results under certain constrains. But when we incorporate geometrical concepts in to thermodynamics, it will further illuminate the hope 
towards the quantization of gravity. Gibbs \cite{gibbs} and Caratheodory \cite{caratheodory} put forward the idea of applying differential geometry in thermodynamics. Later 
Hermann \cite{hermann} and Mrugala \cite{mrugala1,mrugala2} developed the idea of introducing contact geometry in to the thermodynamic phase space. 
In 1976 Weinhold \cite{weinhold} proposed an
alternative approach with a metric, known as Weinhold's metric defined $\textit{ad hoc}$ as the Hessian of the internal energy. The Weinhold's
metric is given by
\begin{equation}
 g_{ij}^{W}=\partial_{i} \partial_{j} U \left(S,N^{r} \right),
 \label{weinholdmetric}
\end{equation}
where $N^{r}$ denotes other extensive variables of the system. Many studies have been done using this metric. But later it is understood that the geometry
based on this metric seems to be physically meaningless in the context of purely equilibrium thermodynamics. In 1979 Ruppeiner \cite{ruppeiner} introduced 
another metric during an attempt to formulate the concept of thermodynamic length.  Ruppeiner's metric is defined as
\begin{equation}
 g_{ij}^{R}=-\partial_{i} \partial_{j} S \left(M,N^{r} \right),
 \label{ruppeinermetric}
\end{equation}
where this metric is conformally equivalent to Weinhold's metric and the geometry that can be obtained from these two are related through a line element
relationship \cite{mrugala3,salamon1},
\begin{equation}
 ds^{2}_{R}=\frac{1}{T} ds^{2}_{W},
 \label{wrconformaltransformation}
\end{equation}
where $T$ denotes the temperature. For systems like ideal classical gas, multicomponent ideal gas, ideal quantum gas, one-dimensional
Ising model, van der Waals model etc, the results obtained with the above two metrics are found to be consistent 
\cite{jany1,jany2,brody1,brody2,brody3,dolan1,dolan2,janke1,janke2,janke3,johnston}. But when we consider black hole 
systems, it is found that these two metrics fail in explaining the properties as well as they lead to many puzzling situations. Among these
inconsistencies, the dependence of metric on the thermodynamic potential is the main problem \cite{salamon2,mrugala4}.
Geometrothermodynamics(GTD) \cite{quevedo1,quevedo2,quevedo3} is the newest approach among the geometric methods. The puzzling properties occurred in the previous methods is due to
the fact that, system possesses different properties, when different thermodynamic potentials are used. But in the frame work of GTD, the metric we are 
considering is invariant with respect to Legendre transformation, hence they are independent of the choice thermodynamic potential of the
system.

To investigate the mathematical structure of thermodynamics, it is necessary to use contact geometry. In GTD, to introduce the
language of differential geometry in thermodynamics, we will consider $(2n+1)$ dimensional thermodynamic phase space $\mathcal{T}$. The 
coordinates of this phase space is defined by the set where $Z^{A}=\{ \Phi, E^a, I^a \}$, where $\Phi$ is the thermodynamic potential, $E^a$ represents
a set of $n$ extensive variables and $I^a$ is the corresponding dual intensive variables, with $a=1,2,.....,n$. Now the contact one form can be written
as
\begin{equation}
 \Theta = d\Phi- \delta_{ab} I^a dE^b ; ~~~~~~~~~~~~\delta_{ab}= diag(1,1,....1)
 \label{oneform}
\end{equation}

The pair $(\mathcal{T},\Theta)$ defines a contact manifold \cite{hermann} if $\mathcal{T}$ is differentiable and $\Theta$ satisfies the condition
$\Theta \wedge (d \Theta)^{n} \neq 0$. Consider $G$ as a non-degenerate metric on $\mathcal{T}$. Then the set 
$(\mathcal{T},\Theta,G)$ defines a Riemannian contact manifold \cite{hermann,hernandez} or the phase manifold. An $n$ dimensional
Riemannian submanifold $\mathcal{E} \subset \mathcal{T}$ is defined as the equilibrium manifold by a smooth map
$\varphi : \mathcal{E} \rightarrow \mathcal{T}$ which satisfies the pullback condition 
$\varphi^*(\Theta) = 0$. Then the metric induced on this equilibrium manifold $\mathcal{E}$, known as Quevedo metric, plays the same role as 
that of Weinhold's and Ruppeiner's metric. This metric can be written as follows,

\begin{equation}
 G=(d\Phi - \delta_{ab} I^a d E^b)^2 +(\delta_{ab} E^a I^b)(\eta_{cd} d E^c d I^d)
\end{equation}
and
\begin{equation}
  g^Q=\varphi^*(G)=\left(E^{c}\frac{\partial{\Phi}}{\partial{E^{c}}}\right)
\left(\eta_{ab}\delta^{bc}\frac{\partial^{2}\Phi}{\partial {E^{c}}\partial{E^{d}}} dE^a dE^d \right)
\label{quevedo metric}
\end{equation}
with $\eta_{ab}$=diag(-1,1,1,..,1) and this metric is Legendre invariant because of the invariance of the Gibbs one form.

This paper is organized as follows. In section \ref{park usual}, we review the Park solution in Ho\v{r}ava gravity and its usual thermodynamics
in details. In section \ref{park gtd}, different thermodynamic geometry methods including GTD are studied in the case of Park black hole
by detailed analysis of both dS and AdS cases. And paper concludes in section \ref{conclusion} with a discussion
regarding the results obtained from the present work.

\section{Park solution in Ho\v{r}ava gravity and its thermodynamics}
\label{park usual}

Let us consider the ADM decomposition of the metric,
\begin{equation}
 ds^2_{4}=-N^2 c^2 dt^2+ g_{ij} \left(dx^i + N^i dt\right)\left(dx^j+ N^j dt\right) ,
 \label{adm}
\end{equation}
and the IR modified Ho\v{r}ava action can be written as
\begin{eqnarray}
 S &=& \int dt d^{3}x \sqrt{g} N [ \frac{2}{\kappa} \left( K_{ij} K^{ij} -\lambda K^{2} \right)-\frac{\kappa^{2}}{2 \nu^{4}} C_{ij} C^{ij}  \nonumber \\
&+& \frac{\kappa^{2}\mu}{2 \nu^{2}} \epsilon^{ijk} R^{(3)}_{il} \nabla_{j} R^{(3)l}_{k} 
 -\frac{\kappa^2\mu^2}{8} R^{(3)}_{ij} R^{(3)ij}+ \frac{\kappa^2 \mu^2 \omega}{8(3\lambda-1)} R^{(3)} \nonumber \\
 &+& \frac{\kappa^{2} \mu^{2}}{8(3\lambda-1)} \left(\frac{4\lambda-1}{4}(R^{(3)})^2-\Lambda_W R^{(3)}+3 \Lambda_W^{2} \right)],
 \label{action}
\end{eqnarray}
where $K_{ij}$ and $C^{ij}$ are the extrinsic curvature and the Cotton tensor, respectively and  
$\kappa,\nu,\mu,\lambda,\Lambda_{W}, \omega$ are constant parameters. 
Among them $\Lambda_{W}$ is related to the cosmological constant by the relation,
\begin{equation}
 \Lambda_{W}=\frac{3}{2} \Lambda.
\end{equation}

The last term in (\ref{action}) represents a soft violation of 
the detailed balance condition \cite{horava1}.
For static and spherically symmetric solution, substituting the metric ansatz as 
\begin{equation}
 ds^2=-N(r)^2 c^2 dt^2+\frac{dr^2}{f(r)}+r^2 \left( d\theta^2+ \sin^2 \theta d\phi^2 \right) ,
 \label{line element}
\end{equation}
in the action (\ref{action}) and after angular integration, we obtain the Lagrangian as

\begin{eqnarray}
 {\mathcal{L}} &=& \frac{\kappa^2\mu^2}{8(1-3\lambda)} \frac{N}{\sqrt{f}} [ (2\lambda-1)\frac{(f-1)^2}{r^2}
 -2\lambda \frac{f-1} {r}f' \nonumber \\  
 &+& \frac{\lambda-1}{2}f'^2 - 2 (\omega-\Lambda_W) (1-f-rf') - 3 \Lambda_W^2 r^2 ] .
\label{lagrangain}
\end{eqnarray}
 Kehagias and Sfetsos \cite{KS} obtained only the asymptotically flat solution (with $\Lambda_W=0$) while 
 Mu-In Park \cite{Park} considered an arbitrary $\Lambda_W$ and obtained a general solution.
 Now the variations with respect to $N$ and $f$ give the equations of motion
 \begin{equation}
(2\lambda-1)\frac{(f-1)^2}{r^2} - 2 \lambda \frac{f-1}{r}f' + \frac{\lambda-1}{2} f'^{2}  2 (\omega- \Lambda_W) (1-f-rf')- 3 \Lambda_W^{2} r^2 =0 ,
\label{equation of motion1}
 \end{equation}
and
 \begin{equation}
 \left( \frac{N}{\sqrt{f}} \right)' \left( (\lambda-1) f'- 2\lambda \frac{f-1}{r} + 2(\omega-\Lambda_W) r \right)
 +(\lambda-1) \frac{N}{\sqrt{f}} \left( f''- \frac{2(f-1)}{r^2} \right)=0 .
 \end{equation}
 
By giving $\lambda =1$ and solving the field equations, we arrive at the Park solution \cite{Park},
\begin{equation}
 N^2=f_{Park}=1+(\omega-\Lambda_W) r^2-\sqrt{r[\omega (\omega-2 \Lambda_W) r^3 + \beta]} ,
 \label{parksolution}
\end{equation}
where $\beta$ is an integration constant related to the black hole mass. Park's solution can easily be reduced to 
 L\"{u}, Mei, and Pope (LMP)'s solution \cite{LMP} as well as Kehagias and Sfetsos (KS)'s solution \cite{KS}.
 
Now let us consider (\ref{parksolution}) in details.  For $r \gg [\beta /|\omega ( \omega-2 \Lambda_W)|]^{1/3}$, we can arrive at two solutions.
First one is the asymptotically AdS case with $\Lambda_W~\textless~0$ and $\omega~\textgreater~0$,
\begin{equation}
f=1+\frac{\left|\Lambda_W \right|}{2} \left| \frac{\Lambda_W}{ \omega} \right| r^2 - \frac{2 M}{\sqrt{1+2 |\Lambda_W/\omega|}} \frac{1}{r} 
+ {\mathcal O}(r^{-4}),
\label{adsfofr}
\end{equation}
and the second one is the asymptotically dS case with  $\Lambda_W~\textgreater~0$ and $\omega~\textless~0$,
 \begin{equation}
f=1-\frac{\Lambda_W}{2} \left| \frac{\Lambda_W}{ \omega} \right| r^2 - \frac{2 M}{\sqrt{1+2 |\Lambda_W/\omega|}} \frac{1}{r} + {\mathcal O}(r^{-4}) .
\label{dsfofr}
\end{equation}
Thermodynamics of Park black hole has been studied in \cite{Park,js2}. Now we will further investigate the different behaviors of these potentials.
In general Park black hole solution has two horizons, one cosmological horizon and the other black hole horizon. By considering the black
hole horizon $r_{+}$, mass of the Park black hole can be written as,
\begin{equation}
M=\frac{1+ 2 (\omega -\Lambda_W) r_+^2 +\Lambda_W^2 r_{+}^{4}}{4 \omega r_{+}}.
\label{mass}
\end{equation}
Using the Bekenstein-Hawking area law,
\begin{equation}
 S=\frac{A}{4}=\pi r_{+}^{2} ,
 \label{bharealaw}
\end{equation}
and the relation 
\begin{equation}
 \Lambda=\frac{(N-1)(N-2)}{2 l^{2}},
\end{equation}
which connects the radius of curvature $l$ of dS or AdS space with $\Lambda$ the cosmological constant (where N is the dimension), 
one can arrive at the mass-entropy relation,
\begin{equation}
 M=\frac{4 S^{2}-4 l^{2} \pi S + l^{4} \pi (\pi+2S\omega)}{4 l^{4} \pi^{\frac{3}{2}} \omega \sqrt{S}}.
 \label{massentropy}
\end{equation}
Particularly in the dS case, also there exists an upper mass bound given by
\begin{equation}
 M_{bound}=\frac{(\frac{4}{l^2} - \omega)}{4} \left( \frac{S}{\pi} \right) ^{3/2}. \label{massbound}
\end{equation}
Thermodynamics regarding this upper mass bound is studied in \cite{js2}.

\begin{figure}
\centering
\resizebox{0.65\textwidth}{!}{
\includegraphics{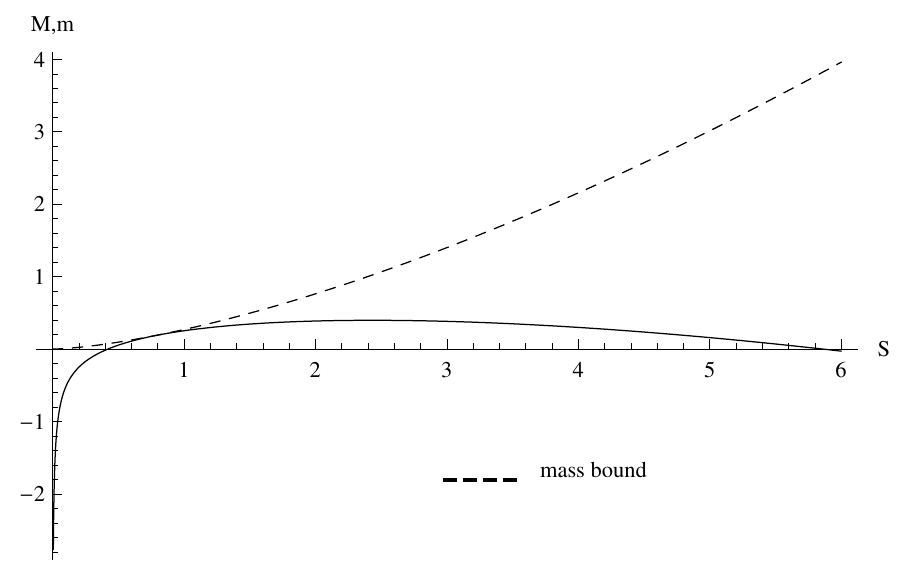}
}
\caption{Plots of mass M vs. entropy for the dS black hole with $l=1$ and $\omega=-2$.}
\label{figmassds}      % Give a unique label
\end{figure}

Now other thermodynamic quantities like temperature, heat capacity and free energy can be obtained from the usual definitions of them,
\begin{eqnarray}
T&=&\left( \frac{\partial M}{\partial S} \right),  \nonumber   \\
C&=&T \left( \frac{\partial S}{\partial T} \right), \nonumber \\
F&=&M- T~S.
\end{eqnarray}

\begin{figure}
\centering
\resizebox{0.65\textwidth}{!}{
\includegraphics{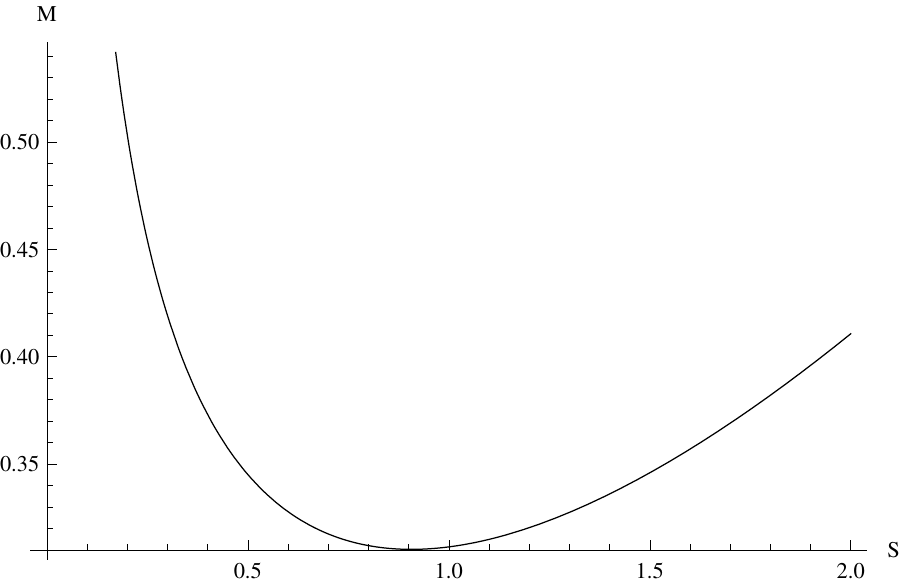}
}
\caption{Plots of mass M vs. entropy for the AdS black hole with $l=-1$ and $\omega=2$.}
\label{figmassads}      % Give a unique label
\end{figure}

\begin{figure}
 \centering
\resizebox{0.65\textwidth}{!}{
\includegraphics{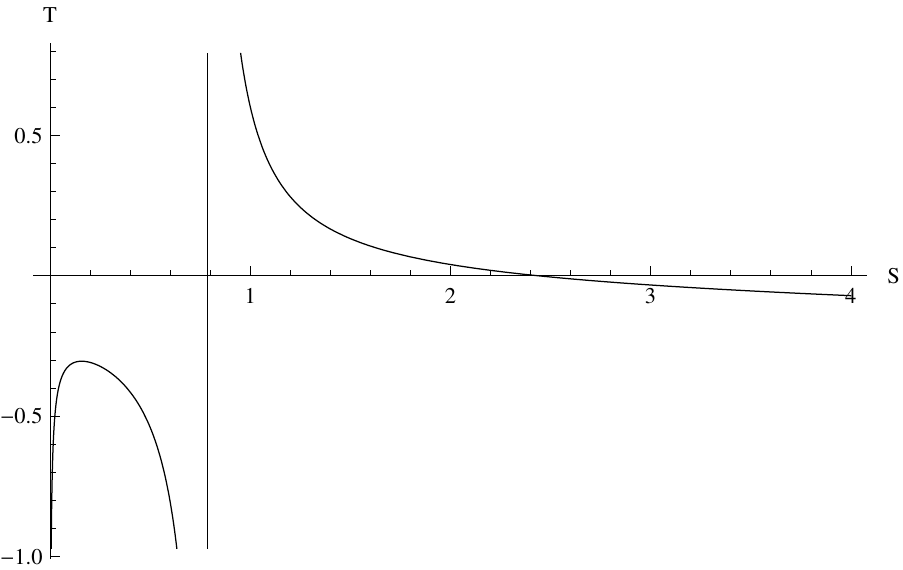}
}
\caption{Plots of temperature T vs. entropy for the dS black hole with $l=1$ and $\omega=-2$.}
\label{figtempds}      % Give a unique label
\end{figure}

\begin{figure}
\centering
\resizebox{0.65\textwidth}{!}{
\includegraphics{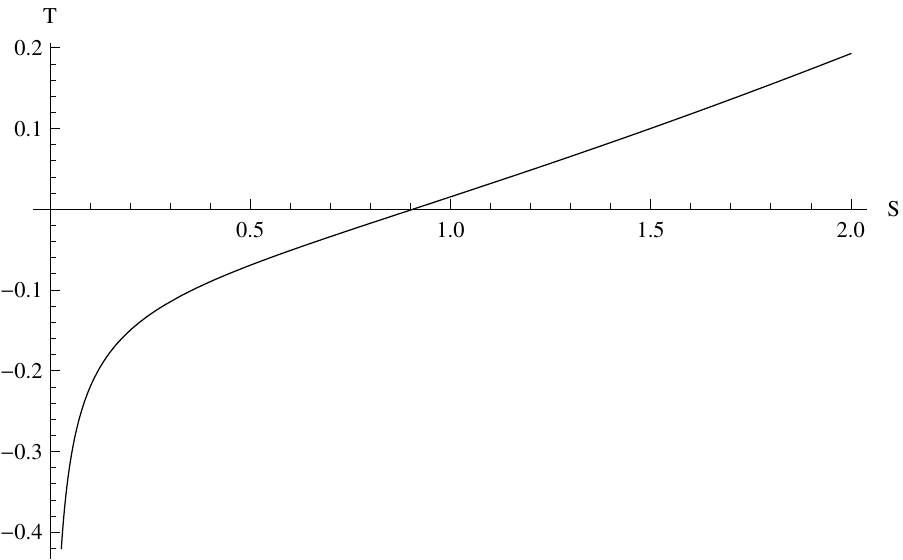}
}
\caption{Plots of temperature T vs. entropy for the AdS black hole with $l=-1$ and $\omega=2$.}
\label{figtempads}      % Give a unique label
\end{figure}

Here the temperature of the black hole is obtained as
\begin{equation}
 T=\frac{12 S^{2}-4 l^{2} \pi S + l^{4} \pi (\pi-2S\omega)}{8 l^{2} \pi^{\frac{3}{2}} \sqrt{S}(l^{2}(\pi+S\omega)-2S)},
\label{temperature}
 \end{equation}
 the heat capacity as,
 \begin{equation}
  C=\frac{2S(12 S^{2}-4 l^{2} \pi S + l^{4} \pi (\pi-2S\omega)(l^{2}(\pi+S\omega)-2S)} 
  {-24 S^3 +4 l^{2} S^2 (7\pi+3S\omega)+ 2 l^{4} \pi S (-5\pi+4S\omega)+ l^{6}\pi(\pi^2+5\pi S \omega -2 S^2 \omega^2)  },
 \label{heatcapacity}
 \end{equation}
and the free energy as,
 \begin{equation}
  F=\frac{-16 S^3 -4 l^{2} S^2 (-6\pi+S\omega)- 12 l^{4} \pi S (\pi+S\omega)+ l^{6}\pi(2\pi^2+7\pi S \omega+2 S^2 \omega^2)}
  {8 l^{4} \pi^{\frac{3}{2}} \sqrt{S} \omega (l^{2}(\pi+S\omega)-2S)}.
  \label{freeenergy}
 \end{equation}
 We have plotted the variation of mass against the entropy in Figs.\ref{figmassds} and \ref{figmassads} for dS and AdS case respectively.
 Similarly, in Figs.\ref{figtempds} and \ref{figtempads}
 temperature 
 variations are plotted. For the dS case (Fig.\ref{figtempds}), we can see that there is an infinite discontinuity in temperature and for a certain 
 range of $S$ values  temperature becomes negative also, which indicates the existence of some unphysical regions. 
 These two anomalous behaviors are due to the existence of mass bound given by (\ref{massbound}). For the AdS case (Fig.\ref{figtempads}) also there
 exist some unphysical regions. Temperature changes continuously in this case without any discontinuities. 

\begin{figure}
 \centering
\resizebox{0.65\textwidth}{!}{
\includegraphics{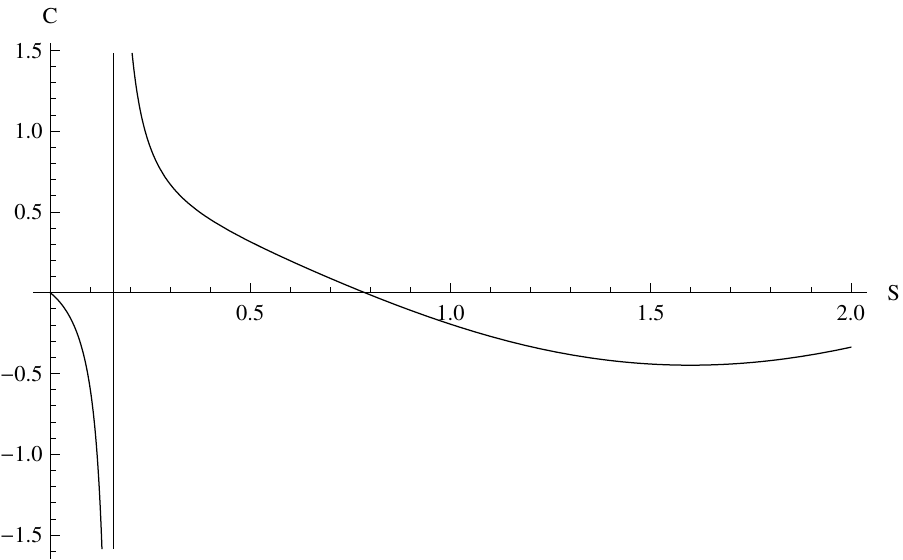}
}
    % Give the correct figure height in cm
\caption{Plots of specific heat C vs. entropy for the dS black hole with $l=1$ and $\omega=-2$.}
\label{figspecds}      % Give a unique label
\end{figure}

\begin{figure}
  \centering
\resizebox{0.65\textwidth}{!}{
\includegraphics{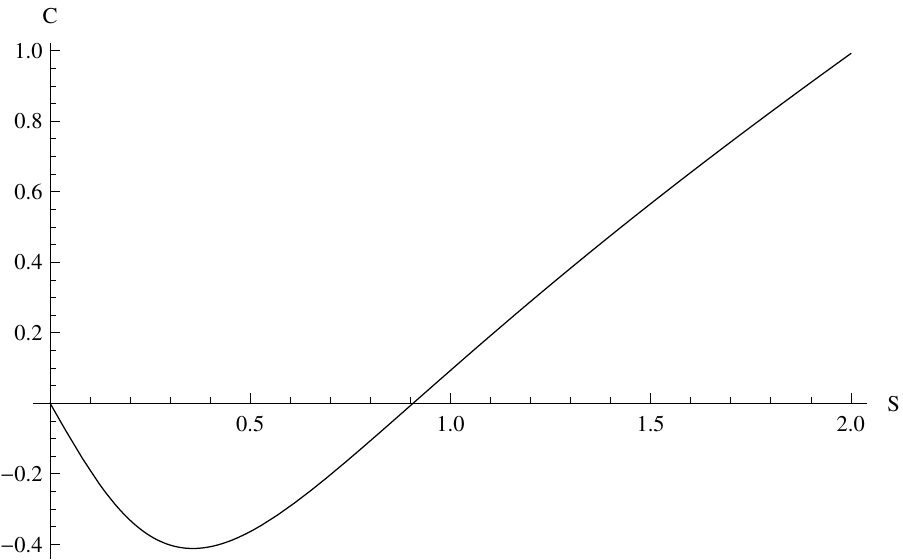}
}
\caption{Plots of specific heat C  vs. entropy for the AdS black hole with $l=-1$ and $\omega=2$.}
\label{figspecads}      % Give a unique label
\end{figure}

\begin{figure}
  \centering
\resizebox{0.65\textwidth}{!}{
\includegraphics{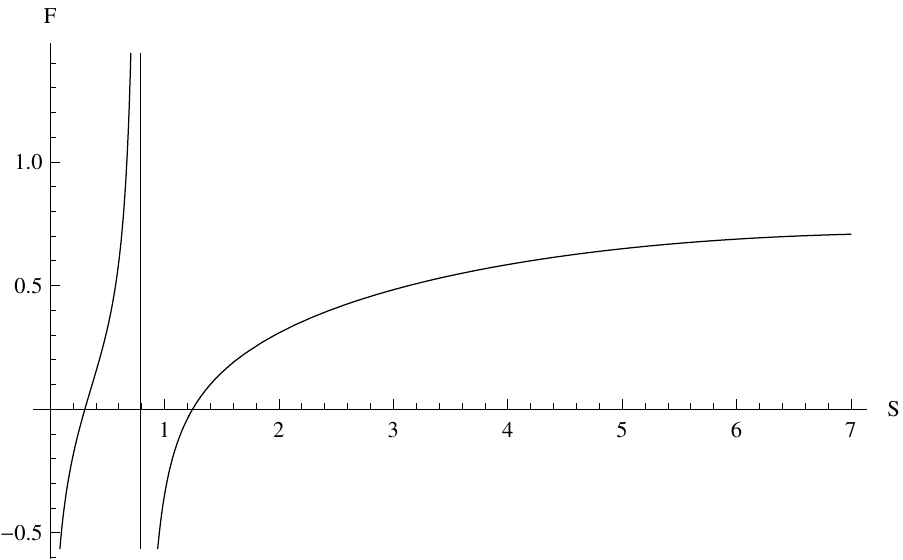}
}
\caption{Plots of free energy F vs. entropy for the dS black hole with $l=1$ and $\omega=-2$.}
\label{figfreeds}      % Give a unique label
\end{figure}

\begin{figure}
\centering
\resizebox{0.65\textwidth}{!}{
\includegraphics{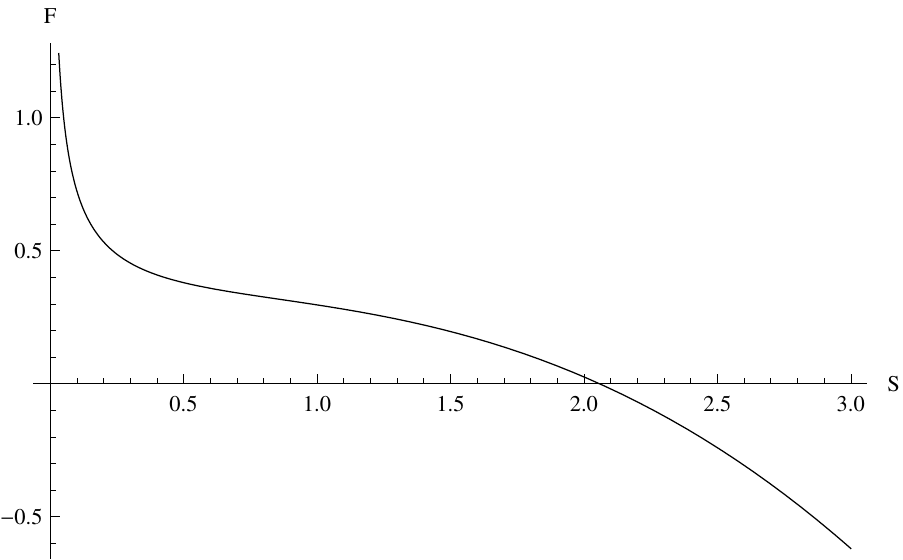}
}
\caption{Plots of free energy F  vs. entropy for the AdS black hole with $l=-1$ and $\omega=2$.}
\label{figfreeads}      % Give a unique label
\end{figure}

In Figs.\ref{figspecds} and \ref{figspecads} we have plotted specific heat of Park black hole with entropy, while in Figs.\ref{figfreeds} and 
\ref{figfreeads}, the variation of free energy against
 entropy is plotted. From Fig.\ref{figspecds} we can see that the Park-dS black hole undergoes a phase transition from thermodynamically unstable
 state to a thermodynamically stable state. In Fig.\ref{figfreeds}, free energy changes from positive to negative, supportingly the black hole changes
 from unstable to stable state via phase transition. But for Park-AdS black hole, from Fig.\ref{figspecads} and Fig.\ref{figfreeads}, we can
 see that black hole undergoes a continuous transition from initial thermodynamically unstable phase to a stable phase and no phase transition takes
 place.
 
So among Park-dS and Park-AdS black hole, only the dS case shows a phase transition. Also there are many regions of these plots, 
like the negative temperature regions, an upper mass bound, infinite discontinuity etc, whose physical 
meaning are still unrevealed. In the next section we will investigate
further regarding this abnormalities shown by the black hole. We are aiming at a good explanation of these observations in terms of different 
thermodynamic geometric methods.

\section{Thermodynamic geometry of Park black hole}
 \label{park gtd}
We now turn to the thermodynamic geometry of Park black hole. In order to incorporate the differential geometry in to the thermodynamic case we 
will consider $l$ and $\omega$ as the other extensive
variables of the present thermodynamic system. Therefore the Weinhold metric can be written from (\ref{weinholdmetric}) as
 \[
         g^{W}=
            \left[ {\begin{array}{ccc}
             M_{SS} & M_{Sl} & M_{S \omega} \\
             M_{lS} & M_{ll} & M_{l\omega} \\
             M_{\omega S} & M_{\omega l} & M_{\omega \omega} \\
                \end{array} } \right]
        \]
where $M_{S}=\partial M / \partial S$, etc. On calculating the curvature scalar of this metric, we can arrive at
\begin{equation}
 R^{W}= \frac{A(S,l,\omega)}{3 [l^2 \pi - 4 S]^3 [8 l^2 \pi S - 36 S^2 + l^4 \pi (5 \pi - 4 \omega S)]^2}.
 \end{equation}
where $A(S,l,\omega)$ is a complicated expression with no physical interest. From the above expression, $R^{W}$ diverges at the points
$S=0.785$ , $S=2.06$ for dS case and at $S=1.171$ for AdS case.(From here, through out this paper we are choosing $l=1$ and $\omega=-2$ for dS
case and $l=-1$ and $\omega=2$ for the AdS case. Also we are not considering imaginary as well as negative roots.).
The point $S=0.785$ or $r_{+}=0.5$ corresponds to the infinite discontinuity of temperature and free energy, and one of the points at which
specific heat becomes zero. Moreover the mass bound is saturated at this point. But Weinhold's metric fails to explain any physical 
singularities in the AdS case.

Now we will consider the Ruppeiner geometry. The Ruppeiner metric can be written from (\ref{ruppeinermetric}) as
\[
         g^{R}=\frac{1}{T}
            \left[ {\begin{array}{ccc}
             M_{SS} & M_{Sl} & M_{S \omega} \\
             M_{lS} & M_{ll} & M_{l\omega} \\
             M_{\omega S} & M_{\omega l} & M_{\omega \omega} \\
                \end{array} } \right].
        \]
The curvature of this metric is given by,
\begin{equation}
  R^{R}= \frac{B(S,l,\omega)}
  {[l^2 \pi - 4 S]^3 [8 l^2 \pi S - 36 S^2 +  l^4 \pi (5 \pi - 4 \omega S)]^2 [4 l^2 \pi S - 12 S^2 + 
    l^4 \pi (\pi - 2 \omega S)][-2 S + l^2 (\pi + \omega S)]^3}.
\end{equation}
where $B(S,l,\omega)$ is also a long complicated expression with less physical interest. For dS and AdS cases, $R^{R}$ possesses singularities 
at points $S=0.785,2.43$ and $S=0.906$ respectively. The point, $S=0.785$ is well explained by Weinhold's metric. But the point, $S=2.43$ or
$r_{+}=0.879$ is the new one that corresponds to zero value of temperature and specific heat. For the AdS case, the point
$S=0.906$ or $r_{+}=0.537$ corresponds to the zeros in mass, temperature and specific heat.

As we mentioned in the introduction, the main problem with Weinhold's and Ruppeiner's matrics is that they are not Legendre invariant. Hence we will
consider the geometrothermodynamics to explain the thermodynamics, since Legendre invariance is preserved in GTD.

%From (**) the easiest way to achieve the Legendre invariance for Weinhold's metric 
%is by applying a conformal transformation with thermodynamic potential as the conformal factor.
%Hence Legendre invariant generalization of $g^{W}$ can be written as
 %\[
  %      g=M g^{W} =M
   %         \left[ {\begin{array}{ccc}
    %         M_{SS} & M_{Sl} & M_{S \omega} \\
     %        M_{lS} & M_{ll} & M_{l\omega} \\
      %       M_{\omega S} & M_{\omega l} & M_{\omega \omega} \\
       %         \end{array} } \right]
        %\]

\begin{figure}
\centering
\resizebox{0.65\textwidth}{!}{
\includegraphics{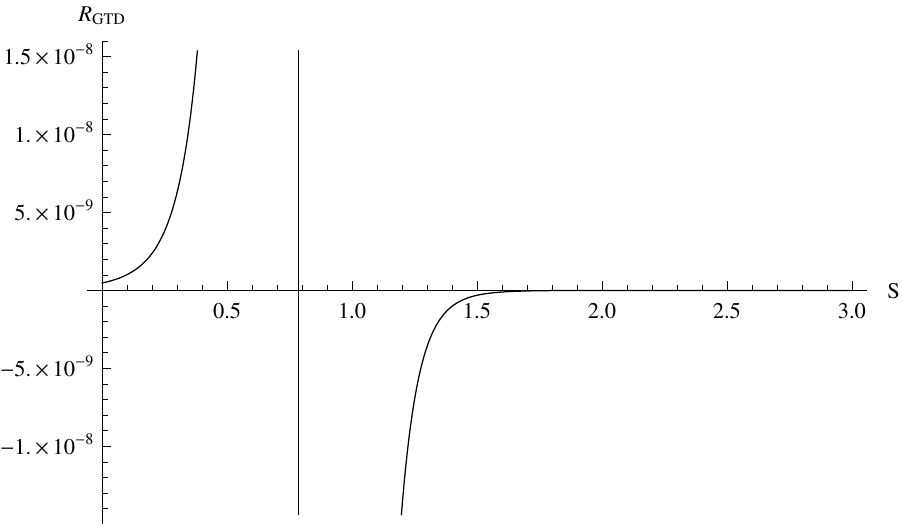}
}
\caption{Plots scalar curvature vs. entropy for the dS black hole with $l=1$, $\omega=-2$}
\label{figdsscalar}      % Give a unique label
\end{figure}

 For Geometrothermodynamic calculations, we will consider 7-dimensional thermodynamic phase space $\mathcal{T}$. This phase space is constituted by the 
 coordinates $Z^A= \{ M,S,l,\omega,T,\iota,\vartheta \}$, where $S,l,\omega$ are extensive variables and $T,\iota,\vartheta$ are their dual intensive
 variables. Then the fundamental Gibbs 1-form defined on $\mathcal{T}$ can be written as,
 \begin{equation}
  \Theta=dM-TdS-\iota dl-\vartheta d\omega
 \end{equation}
 The equilibrium phase space $\mathcal{E}$ can be defined as a simple maping
 $\varphi: \{S,l,\omega \} \rightarrow \{ M(S,l,\omega), S, l, \omega, T(S,l,\omega),\iota(S,l,\omega),\vartheta (S,l,\omega)$. The 
 Quevedo metric is given from (\ref{quevedo metric}),
 \[
         g^{GTD}=(SM_{S}+lM_{l}+\omega M_{\omega})
            \left[ {\begin{array}{ccc}
             -M_{SS} & 0 & 0 \\
             0 & M_{ll} & M_{l\omega} \\
             0 & M_{\omega l} & M_{\omega \omega} \\
                \end{array} } \right].
        \]
The curvature scalar corresponding to the above metric is found to be,
\begin{equation}
  R^{GTD}= \frac{D(S,l,\omega)}
  {[l^2 \pi - 2 s]^3 [4 l^2 \pi s + 12 s^2 + l^4 \pi (3 \pi - 2 \omega s)]^2 [-20 l^2 \pi s + 28 s^2 + 
    l^4 \pi (3 \pi - 2 \omega s)]^3}.
\end{equation}
in which $D(S,l,\omega)$ is a complicated expression with less physical interest. At points $S=0.785$ and at $S=0.477$ and $2.1$ for dS and AdS respectively, 
the Legendre invariant scalar curvature becomes zero or shows infinite discontinuities. The point $S=0.785$ or $r_{+}=0.5$ is the same point where 
the phase transition takes place. To get an exact idea regarding this, we will consider the Fig. \ref{figdsscalar}, which shows the
correspondence between the divergence of scalar curvature $R^{GTD}$ and specific heat $C$. It is very interesting to note that the point
$S=0.477$ or $r_{+}=0.386$ in AdS case corresponds to the point of inflection in the curves of temperature, specific heat and free energy, where the 
convex nature of curve changes to concave nature or vice versa. Similarly the point $S=2.1$ or $r_{+}=0.817$ coincides with the point of free energy curve
where it becomes zero. So using geometrothermodynamics and hence by constructing the Legendre invariant metric, we are able to reproduce the behavior
of thermodynamic potentials and their interactions. The correspondence of divergence and zeros of thermodynamic potentials with 
the divergence of Legendre invariant scalar curvature leads to the complete understanding of Park black hole thermodynamics.

\section{Conclusion and discussion}
\label{conclusion}

In this paper, we have investigated the thermodynamics as well as thermodynamic geometry of Park black hole. We have considered both dS and AdS cases. 
We have analyzed the usual thermodynamics of both these cases and found that there exist many abnormal behaviors like existence
of an upper mass bound, negative temperature, infinite discontinuity in temperature, heat capacity and free energy, etc. 
We have incorporated the geometric ideas in to the usual thermodynamics by means of different thermodynamic geometric methods. 

We have analyzed first the thermodynamic geometry based on Weinhold's metric and Ruppeiner's metric and the GTD. We have found that the corresponding 
thermodynamic scalar curvature possesses many singularities, and these singularities are in accordance with the behaviors of mass, temperature, 
specific heat and free energy. As we have mentioned in this work, these two methods depend entirely on the choice of thermodynamic
potential to describe the system. Even though this particular choice gives almost good results, but the lack of Legendre invariance leads us to
consider much more general geometrothermodynamic method. The potential independence of the results or in other words the Legendre invariance is 
assured in this metric.

When we use GTD to explain the thermodynamics, we find that it possesses a true curvature singularity. And the singularity corresponds to the
points where the mass bound gets saturated, temperature shows infinite discontinuity and specific heat also
shows infinite discontinuity. Park dS black holes undergo a second order phase transition from a thermodynamically
unstable state to thermodynamically stable state while in the AdS case, there exists no such behaviors. So GTD
reproduces the thermodynamics of Park black hole irrespective of the potential choice to explain the system. When we consider the 
GTD metric, it is found to be finite and smooth at the regions where the black hole is stable. But when black hole becomes unstable, this metric 
possesses true singularities, and as mentioned above, this corresponds to the second order phase transition shown by the black hole.
So by incorporating the Legendre invariance as well as differential geometry, GTD is an important method to well explain the 
thermodynamics of black holes. Here GTD explains the second order phase transition, existence of negative temperature, point of inflection and
the upper mass bound of Park black hole.

\section*{Acknowledgements}
The authors wish to thank UGC, New Delhi for financial
support through a major research project sanctioned to VCK. VCK also
wishes to acknowledge Associateship of IUCAA, Pune, India.

\end{document}